\documentclass[aps, pre,10pt,notitlepage,nofootinbib,superscriptaddress,showkeys,twocolumn]{revtex4-1}
\usepackage{amstext,amsmath,amssymb,amsfonts,bbm, amsthm}
\usepackage[latin1]{inputenc}
\usepackage{fancyhdr}
\usepackage{graphicx}
\usepackage{hyperref}
\usepackage{tocvsec2}

\usepackage{color}

\def\beq{\begin{equation}}
\def\be{\begin{equation}}
\def\ee{\end{equation}}
\def\bes{\begin{eqnarray}}
\def\ees{\end{eqnarray}}

\def\f{\frac}

\def\pp{\partial}

\begin{document}

\title{\large \bf Self-organization without heat: the geometric ratchet effect}

\author{{Matteo Smerlak}}\email{smerlak@aei.mpg.de}
\affiliation{Max-Planck-Institut f\"ur Gravitationsphysik, Am M\"uhlenberg 1, 14476 Golm, Germany}

\author{{Ahmed Youssef}}\email{youssef@mathematik.hu-berlin.de}
\affiliation{Institut f\"ur Physik, Humboldt-Universit\"at zu Berlin, Rudower Chaussee 25, 12489 Berlin, Germany}

\date{\small\today}

\begin{abstract}\noindent
We point out a surprising feature of diffusion in inhomogeneous media: under suitable conditions, the rectification of the Brownian paths by a diffusivity gradient can result in initially spread tracers spontaneously \emph{concentrating}. This ``geometric ratchet effect'' demonstrates that, in violation of the classical statements of the second law of (non-equilibrium) thermodynamics, self-organization can take place in thermodynamic systems at local equilibrium without heat being produced or exchanged with the environment. We stress the r\^{o}le of Bayesian priors in a suitable reformulation of the second law accommodating this geometric ratchet effect.

\end{abstract}
\maketitle

\section{Introduction}

Can \emph{isolated} thermodynamic systems self-organize? To this question, most physicists would answer in the negative, because ``the second law forbids it''. Moreover, if asked to provide an example of a physical process precluding self-organization, many would mention \emph{diffusion}. The purpose of this paper is to show that not only can isolated thermodynamic systems self-organize, but also that one driving mechanism for this phenomenon is diffusion itself. We argue furthermore that, although it does not violate the second law of thermostatics, this effect does provide a counter-example to various classical statements to the effect that ``the entropy of an isolated system can never decrease'', including the Clausius-Duhem inequality of linear irreversible thermodynamics \cite{DeGroot1984}. 

Our model system is simple and has been studied many times: a gas of non-interacting tracers diffusing within a structurally inhomogeneous medium. By ``structurally'', we mean that the inhomogeneities are not forced into the systems by an external source of energy (a laser, heated boundaries, an external stress, etc.), but are frozen within the medium itself. The typical example we have in mind is an inhomogeneous fluid or solid mixture with very slow relaxation dynamics. We will give a concrete example of such a system in the next section. 

Using the Fokker-Planck diffusion equation \cite{VanKampen1992}, we show that such inhomogeneous media can rectify the random motion of tracers in such a way that the normal effect of diffusion is \emph{reversed}: instead of spreading to reach a homogeneous equilibrium, the tracers of our setup tend to concentrate in the region of space where their diffusivity is lowest. This phenomenon takes place on macroscopic, but finite time scales---essentially until the structural inhomogeneities get smoothed out by their own diffusion process.  

This surprising effect is very analogous to the well-known \emph{ratchet phenomenon} \cite{Reimann2002,Gabrys2004}, whereby the fluctuations of Brownian particles are rectified by a periodic potential with broken spatial inversion symmetry, resulting in directed transport of the particles. The essential difference between this effect and the one described in this paper is the fact that, to have a non-zero net effect on the diffusive motion of the particles, conventional ratchets must be kept away from equilibrium by an external forcing, often temperature oscillations. This forcing involves \emph{heat exchanges} between the ratchets and their environment, a situation that we exclude from our considerations. To highlight this difference, we propose to call the effect here described the ``geometric ratchet effect''.\footnote{We understand the word ``geometry'' as referring to the material ``structure of space'': here the inhomogeneous diffusivity of the medium.}

The physics underlying the geometric ratchet effect is the \emph{separation of two timescales}: the one associated to the diffusion of tracers, and the one associated to the relaxation of the medium itself. This circumstance is what allows the system to depart from the behavior expected for thermodynamic systems in local equilibrium and lead to spontaneous self-organization. We will see that it is possible to reformulate the second law of (non-equilibrium) thermodynamics that is applicable in this context by using a suitable \emph{relative entropy} \cite{Kullback1959}.

Our paper is organized as follows. In sec. \ref{indiff}, we introduce the physical mechanism underlying the geometric ratchet effect, namely inhomogeneous (or ``state-dependent'') diffusion. We then spell out in sec. \ref{antidiffusion} the conditions for self-organization of the tracers, and show that these conditions can easily be met, for instance with fluid mixtures. In sec. \ref{entropybalance}, we address the fate of the second law in this situation, and consider in sec. \ref{discussion} some of the questions which may come to mind concerning our prediction. Sec. \ref{conclusion} contains our conclusions.


\section{Inhomogeneous diffusion}\label{indiff}

Diffusion is the most basic and universal relaxation mechanism. Following directly from the central limit theorem, it arises in any situation where the path of the relevant tracers (particles, defects, light waves, etc.) consists of a large number of independent and identically distributed\footnote{The step probability distributions should have finite mean and variance; other macroscopic equations arise when this condition is dropped, e.g. with L\'evy flights.} random steps. As understood by Einstein \cite{Einstein1905} and Smoluchowski \cite{Smoluchowski1906}, this results in the continuum limit in the diffusion equation for the tracers probability density $p(x,t)$,
\be\label{diffusion}
\f{\pp p}{\pp t}=D\Delta p,
\ee 
where $D$ is the (positive) diffusivity. As a rule, this behavior is accompanied by an increase of disorder in the system, expressing the loss of spatial information in the process. If this spatial disorder is measured by the usual \emph{positional entropy} 
\be\label{gibbs}
S_{\textrm{pos}}(t)=-\int dV\, p\ln p,
\ee
where $dV$ is the spatial volume measure, a simple computation using the diffusion equation \eqref{diffusion} gives
\be
\f{dS_{\textrm{pos}}}{dt}=D\int dV\, \f{(\nabla p)^{2}}{p}\geq0.
\ee
This confirms that, in normal circumstances, diffusion can only increase the disorder in a system. Anticipating on our discussion in sec. \ref{entropybalance}, let us stress that the above inequality is not \emph{per se} a statement of the second law of thermodynamics, even when it leads to the same conclusion (disorder increases): the second law is concerned with \emph{thermodynamic} entropies, which have contributions from the dispersion of the particles in velocity space as well as in physical space. 

Now, one key assumption underlying the derivation of the diffusion equation \eqref{diffusion} is the \emph{homogeneity} of the medium. This is necessary for the ``identically distributed'' condition which allows the central limit theorem to apply. Without this assumption, it is not clear \emph{a priori} what the equation for the probability density $p$ should be. In particular, if $D=D(x)$ is a space-dependent function because of slowly relaxing temperature or mobility gradients (or both), one could envisage (among other possibilities) a generalized diffusion equation with Fokker-Planck (FP) form
\be\label{FP}
\f{\pp p}{\pp t}=\Delta(Dp),
\ee 
or with Fourier-Fick (FF) form
\be\label{FF}
\f{\pp p}{\pp t}=\nabla\cdot(D\nabla p).
\ee
Attempts to find the correct form for the diffusion equation in inhomogeneous media have led to a long-lasting discussion in the literature, see e.g. \cite{VanKampen1988,VanMilligen2005,Sattin2008,Bringuier2011}. Its conclusion is that, in van Kampen's words, ``no universal form of the diffusion equation exists, but each system has to be studied individually'' \cite{VanKampen1988}. 

\begin{figure}[t]
\includegraphics[scale=.38]{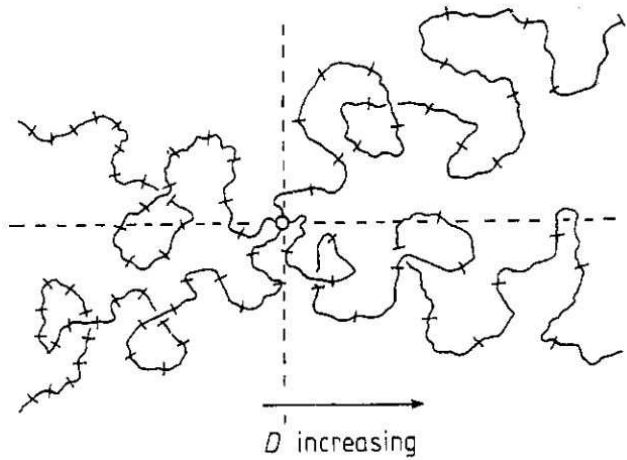}
\caption{Typical Brownian paths for FP-type systems, illustrating their ``inhomogeneous but isotropic'' character; reproduced from \cite{Collins1989}.}
\label{joli}
\end{figure}

This notwithstanding, it is possible to get a general intuition about the respective applicability of equations \eqref{FP} and \eqref{FF} by means of microscopic random walks models \cite{Collins1989,Collins1997,Bringuier2011}. Since in the following we shall focus on the FP equation, let us briefly describe one such model leading to \eqref{FP} in the continuum limit \cite{Collins1997,Bringuier2011}. Consider a (say one-dimensional) lattice with spacing $a$, where a particle can hop from each site to its nearest neighbors with equal probability $1/2$, and assume that the hopping rates $W_{n}$ depends on the sites $n$. Then the particle population $N_{n}$ evolves according to
\be
\f{dN_{n}}{dt}=-W_{n}N_{n}+\f{1}{2}W_{n+1}N_{n+1}+\f{1}{2}W_{n-1}N_{n-1}.
\ee
Using standard manipulations, it is easy to see that in the $a\rightarrow0$ limit the normalized density $n(x=na,t)=N_{n}/\sum_{k}N_{k} a$ satisfies the FP equation \eqref{FP} with $D(x)=a^{2}W(x)/2$. (A more sophisticated microscopic model of the FP equation, based on a random Lorentz gas with varying free volume fraction, has been discussed recently in \cite{Tupper}.\footnote{In fact, \cite{Tupper} shows that the random Lorentz gas model can lead to Fourier-Fick diffusion equations as well when the spheres' radii vary at constant free volume fraction. This paper also comments on the tension between the FP equation and the principles of statistical mechanics, with different conclusions from ours.})

The lesson we learn from this simple random walk model is that the FP equation arises whenever the microscopic dynamics is inhomogeneous but \emph{isotropic}, viz. does not favor hops in one particular direction \cite{Collins1989}; the presence of such a microscopic bias, due for instance to the vicinity of walls, results in either the FF equation \cite{Lancon2001,Lau2007} or some hybrid equation. An pictorial view of typical Brownian paths for FP-type systems is provided in Fig.


Let us emphasize that FP-type systems are not theoretical abstractions: the recent literature reports several observations of inhomogeneous diffusion following the FP equation. Among such systems, we can mention fusion plasmas \cite{VanMilligen2005a} or thermodiffusion at certain values of the Soret coefficient \cite{Collins1989}. 

\begin{figure}[t]
\includegraphics[scale=.45]{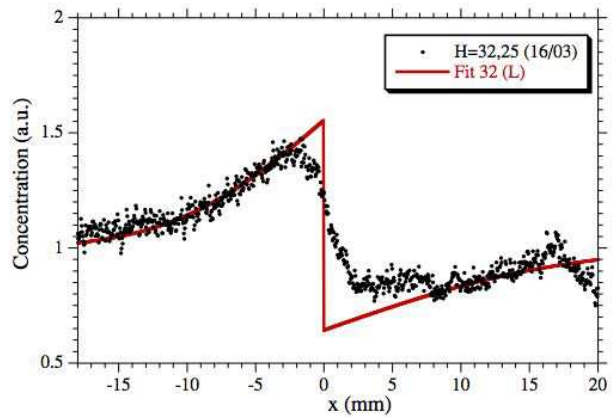}
\caption{The results of \cite{VanMilligen2005} for the diffusion of food coloring in an inhomogeneous water-gelatine mixture: the dots are measured values of the tracer concentration after $32$ hours of diffusion (with a homogeneous initial distribution), the solid line is a fit with the analytical solution of the corresponding FP equation.}
\label{milligen}
\end{figure}

For the sake of definiteness, let us describe explicitly one concrete setup exemplifying our concept of ``structural inhomogeneities'' which has been studied experimentally by van Milligen \emph{et al} \cite{VanMilligen2005}. Prepare two mixtures of water and gelatine $1$ and $2$ with different relative concentrations $n_{1}$ and $n_{2}$. Consider a closed, tube-like vessel filled for the first half with a mixture $1$, and for the second half with mixture $2$, in both of which an equal quantity of food coloring has been dissolved. Because the presence of gelatine in water modifies the viscosity, the fluid medium has inhomogeneous diffusivity, and one can check whether the FP equation or some other equation applies to the diffusion of the food coloring; van Milligen \emph{et al}. proved in  \cite{VanMilligen2005} that the former is the case in their setup. In Fig. \ref{milligen} reproduced from this reference, the measured concentration of coloring across the vessel is compared to a fit with the analytical solution of the FP equation ($x=0$ is the boundary between mixtures $1$ and $2$): the match is very good except close to the boundary, and can actually be improved by considering higher-derivative terms in the underlying master equation (of which the FP equation is an approximation), see \cite{VanMilligen2005}.\footnote{Note that the FF equation would predict instead a \emph{flat} concentration profile, and therefore clearly not applicable in this case.}

\section{Anti-diffusion}\label{antidiffusion}

\begin{figure*}[t]
\includegraphics[scale=1]{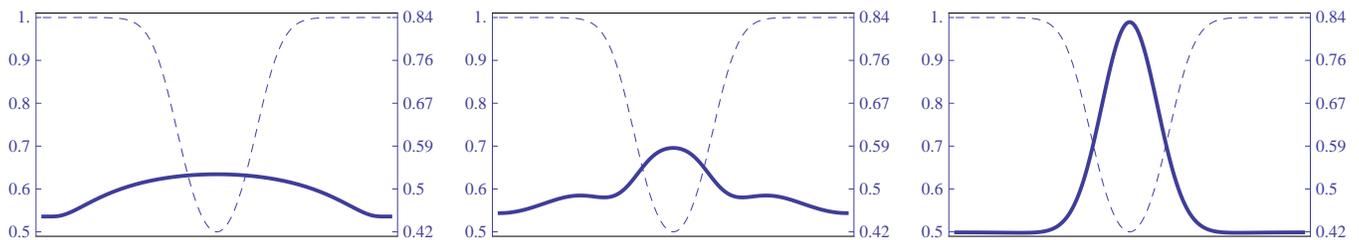}
\caption{Numerical solution of the one-dimensional FP equation with Neumann boundary conditions, at three times $t_{0}<t_{1}<t_{2}$ (left to right). The dashed curve (left axis) is the local diffusivity $D(x)$ normalized to its maximal value; the thick curve (right axis) are the probability densities $p(t_{i},x)$, with $x$ in arbitrary units. The amplitude of variation of $D(x)$ is consistent with the fluid experiment of Ref. \cite{VanMilligen2005}.}
\label{numsol}
\end{figure*}

\begin{figure}[b]
\includegraphics[scale=0.9]{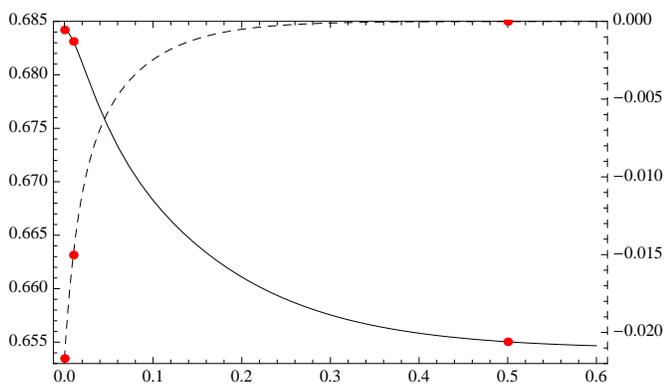}
\caption{The positional entropy (continuous curve, left axis) and relative positional entropy (dashed curve, right axis) in the ``trap'' of Fig. \ref{numsol} as functions of time (arbitrary units). The red dots indicate the times $t_{0}, t_{1},t_{2}$ plotted in Fig. \ref{numsol}.}
\label{entropy}
\end{figure}

The example of the water-gelatine mixture is particularly striking, because it allows one to see with the naked eye the most remarkable consequence of the FP equation: the equilibrium distribution $p^{*}(x)$ is not homogeneous, but rather $p^{*}(x)\propto1/D(x)$ (see Fig. 3 of the aforementioned reference). This striking property (which is sometimes argued to be impossible \cite{Lancon2001,Lau2007}, but was directly observed in \cite{VanMilligen2005}) is precisely what drives the geometric ratchet effect. Hence, from now on we focus exclusively on inhomogeneous media with FP-type diffusive dynamics. 

Does the positional entropy of tracers always increase in this setup? Using the FP equation, we find 
\be\label{entropyvariation}
\f{dS_{\textrm{pos}}}{dt}=\int\, D\,\nabla p\cdot\left(\f{\nabla p}{p}+\f{\nabla D}{D}\right)\,dV.
\ee
Hence, we see that $dS_{\textrm{pos}}/dt<0$ whenever $\nabla p$ and $\nabla D$ have opposite directions and $\nabla D$ has sufficiently large norm, over a sufficiently large region of space. These are the conditions for diffusion to result in spatial self-organization rather than increase of disorder.

A simple example demonstrating this phenomenon can be obtained by constructing a (say spherically symmetric) ``trap'', in which $D=D(r)$ is a monotonously increasing function of some radial coordinate $r$, and taking a bump-like initial distribution $p_{0}(r)$, with $p_{0}(r)$ mildly decreasing with $r$. The numerical solution of the one-dimensional FP equation in this case is plotted in Fig. \ref{numsol}, and the function $S_{\textrm{pos}}(t)$ in Fig. \ref{entropy}: the probability distribution clearly concentrates about $r=0$, thus exhibiting what may be called ``anti-diffusion''. A direct check of this effect could be obtained numerically, e.g. with a random Lorentz gas model \cite{Tupper}, or experimentally, e.g. a with a suitably prepared water-gelatine mixture \cite{VanMilligen2005}.


\section{On the second law}\label{entropybalance}

Does the geometric ratchet effect violate the second law? To answer this question, let us carefully distinguish three---nonequivalent---statements:

\begin{enumerate}
\item
\emph{Second law of thermostatics (Clausius inequality):} if $A$ and $B$ are two equilibrium states of an isolated macroscopic system with thermodynamic entropy $S(A)$ and $S(B)$, and $B$ can be reached from $A$ by an adiabatic transformation, then $$S(B)-S(A)\geq0.$$
\item
\emph{Second law of irreversible thermodynamics (Clausius-Duhem inequality):} the entropy production rate $\sigma(x,t)$ of an isolated continuous medium in local equilibrium with dissipative fluxes $j_{a}(x,t)$ and conjugate thermodynamic forces $X_{a}(x,t)$ satisfies 
$$\sigma=\sum_{a}\,j_{a}\cdot X_{a}\geq0.$$
\item
\emph{Law of entropy increase (folklore):} the entropy of an isolated system can never decrease.
\end{enumerate}
These statements by no means exhaust what is meant by the phrase ``second law'' in the literature (the book \cite{Capek2005} compiles 21 different statements). In particular, it is well-known that the second law of thermostatics (statement $1.$) has several other formulations, involving the notions of exchanged heat or dissipated work.

Now, the second law of thermo\emph{statics} \cite{Balian2003} is clearly not threatened by our setup, which is by nature out of equilibrium. What is at stakes here is the second law of thermo\emph{dynamics} (statement $2$.), namely the Clausius-Duhem (CD) inequality. We now show that it is actually violated by the geometric ratchet effect.


Assume from now on that the diffusivity gradients in the medium are \emph{not} due to temperature variations, but rather to variations of the tracers' mobility. This is the case for instance in the experiment of Ref. \cite{VanMilligen2005} where food coloring diffuses in a fluid with varying viscosity. In this case, no heat transfer accompanies the diffusion of the tracers, because no information leaks into internal degrees of freedom during the process: being in local equilibrium with the medium, their microscopic velocity distribution remains Maxwellian with fixed temperature at all times. Hence, the only dissipative flux is the particle flux
\be\label{flux}
j=-\nabla(Dp)
\ee 
defined by the continuity equation $\pp p/\pp t+\nabla\cdot j=0$. The corresponding thermodynamic force is, according to classical irreversible thermodynamics \cite{DeGroot1984},
\be\label{force}
X=-\nabla\Big(\f{\mu}{T}\Big),
\ee
where $\mu$ is the local chemical potential of the tracers and $T$ is the constant temperature. Since the latter are by assumption non-interacting, the ideal gas formula gives
\be
\f{\mu}{T}=\ln p + \textrm{constant},
\ee
and therefore 
\be
X=-\f{\nabla p}{p}.
\ee 
Dotting \eqref{flux} with \eqref{force} then gives the entropy production rate 
\be
\sigma=\nabla(Dp)\cdot\f{\nabla p}{p},
\ee 
which is readily seen to coincide with the integrand of equation \eqref{entropyvariation}. This confirms that in the absence of heat transfer, the production rate of thermodynamic entropy coincides with the time derivative of the positional entropy $S_{\textrm{pos}}$ of the tracers. In this sense, the decrease of the latter in our setup implies a (transient) violation of the Clausius-Duhem inequality. 

We note that the same conclusion would be reached in the most recent versions of non-equilibrium thermodynamics, such as ``stochastic thermodynamics'' \cite{Seifert2008}, which can also be applied to small systems with relatively large fluctuations, and where ``thermodynamic forces'' cannot be defined. Thus, for an ensemble of tracers diffusing within a medium, Seifert defines a stochastic ``system entropy'' whose ensemble average is given by the classical formula \eqref{gibbs}, and ``entropy production in the medium'' as the work received by the tracers from external forces \cite{Seifert2005} (see also \cite{VandenBroeck2010}). Hence, in the absence of such forces\footnote{One might argue that the variations of diffusivity are effectively equivalent to a ``force term'' proportional to $\nabla D$, and therefore that it is not the case that the tracers do not exchange heat with their environment. This is only half true: if $\nabla D$ does appear to behave as an effective (one is tempted to say ``entropic'') force as far as the motion of the tracers is concerned, we must emphasize that \emph{it does not produce work}.} but with a diffusivity gradient of the kind considered here, the total Seifert entropy decreases, in contradiction with his ``fluctuation theorem'' \cite{Seifert2005}. As the reader will have anticipated, the key assumption underlying this identity, as well as the results of \cite{VandenBroeck2010}, is the homogeneity of the medium, and in particular $D=\textrm{const}$. We see here that no such relation holds in inhomogeneous media---at least not as such.

Let us now come to the law of entropy increase (statement $3.$). As such, this statement is manifestly too vague to be challenged by any actual physical situation, for ``entropy'' (or for that matter ``disorder'') is not a well-defined notion \emph{a priori} \cite{Balian2003}. As Baez puts it, ``whenever you're tempted to talk about entropy, you should talk about relative entropy'' \cite{Baez}. What our example shows is that the positional entropy \emph{relative to the uniform prior} can decrease in a diffusion process, or in other words that the latter can reduce a disorder conceived as ``equiprobability in space''. 

Suppose now that ``disorder'' is defined as ``$p$ equals its equilibrium distribution $p^{*}$''. The corresponding \emph{relative positional entropy} is
\be
S_{\textrm{pos}}(t\,\vert\, p^{*})=-\int\, p\,\ln\f{p}{p^{*}}\,dV,
\ee
and using $p^{*}\propto D^{-1}$, we check that
\be
\f{dS_{\textrm{pos}}(t\,\vert\, p^{*})}{dt}=\int\, \f{\big(\nabla (Dp)\big)^{2}}{Dp}\,dV\geq0.
\ee
Hence, the positional entropy relative to the equilibrium distribution associated to the inhomogeneous medium \emph{does increase}, see also Fig. \ref{entropy}. This result means that the ``distance'' between $p$ and its final state $p^{*}$ can only decrease in time, even when the latter state is actually more ``ordered'' (the tracers are more concentrated) than the former. This, in turn, is the proper formulation of irreversibility in the present context.

Thus, what our observations show is not that the second law of thermodynamics is ``wrong''. Instead, it highlights a condition for its applicability in standard form which had perhaps not been fully appreciated so far: it applies when \emph{all the dynamical variables relax on the same timescale}. If some of them happen to be frozen in an inhomogeneous state\footnote{The reader can think of this situation as ``partial ergodicity breaking'', with some variables (here the tracers) evolving ergodically in the background set by the other, non-ergodic ones (the medium).} (such as the medium's diffusivity distribution here), it \emph{is} possible that self-organization can result without the need for an external energy source and without internal heating; in this case, a consistent formulation of the second law of thermodynamics should take this prior into account. Similar conclusions were reached by Banavar and Maritan in \cite{Banavar} following a line of thought pioneered by Jaynes \cite{Jaynes1957a}.



%
%
%

\section{Discussion}\label{discussion}

To avoid possible misunderstandings, we now list and answer some questions which---we feel---are naturally prompted by our argument.

\paragraph{How is this effect different from diffusion in a (gravitational or electric) field?} Simply by the fact that there is no field in our setup! The concentration of tracers in regions of low diffusivity happens without any external forcing, and thus cannot be compared to a phenomenon such as sedimentation. This is what makes the trap of sec. \ref{antidiffusion} a ratchet instead of a mere potential well. 

\paragraph{The tracers are clearly not isolated, so why call this effect ``self-organization''?} The tracers are not, but the system tracers + medium is. It is this composite system which ``self-organizes'' in the process, and by assumption the latter is isolated.

\paragraph{How is this effect different from gravitational collapse or droplet formation in non-miscible fluids? Isn't that also ``self-organization''?} It is not, because the latter effects are driven by attractive interactions between the tracers, and hence conservation of energy results in the heating of the system. This means that, as the spatial distribution of particles becomes sharper, the velocity distribution becomes spreads out and compensates the superficial ``self-organization'' of the system. Our systems, on the other hand, do not involve any interactions between the tracers, and therefore do not lead to any heating.   

\paragraph{The FP equation \eqref{FP} is only one of many interpretations (Ito, Stratonovitch, Klimontovitch...) of the microscopic Langevin equation. It is not canonical, and other interpretations lead to different equilibrium solutions.} This is not an issue of interpretation: the main physical consequence of the FP equation, namely the fact that the equilibrium distribution is inhomogeneous, has been observed in the lab. Which Langevin equation and interpretation thereof one chooses to model diffusion with, is a matter of personal taste.

\paragraph{Take a glass of water, cool it below the fusion point, and wait---its entropy decreases. There is nothing mysterious about self-organization.} The crystallization of water does not take place unless the latent heat is evacuated in some way, and this is of course not a mystery. The point is that no such heat transfer between the system (tracers + medium) and its environment is involved in the geometric ratchet effect.

\paragraph{You have claimed but not showed that the diffusion of tracers does not involve heat fluxes. These may make the total entropy increase.} The physics involved in our setup is plain particle diffusion; if the latter came with heat fluxes in inhomogeneous media, it would in homogeneous media as well---which is clearly not the case.

\paragraph{Transient violations of the second law are well known, and in fact manifest e.g. in Seifert's fluctuation theorem $\langle e^{-\Delta s}\rangle=1$, so what is new here?} The transient violations of the second law observed in \cite{Wang2002} and quantified by the celebrated fluctuations theorems \cite{Jarzynski2011} are fluctuations of \emph{small} systems, which disappear in the thermodynamic limit. On the contrary, the effect considered here is macroscopic, both in space and in time. 

\paragraph{Violations of the $H$-theorem in real gases are well known. Aren't you confusing the latter with the second law?} We are not. The violations of the $H$-theorem (monotonous increase of the single-particle entropy) first discovered by Jaynes \cite{Jaynes1971} are due to the interactions between the molecules of a real gas, which are not taken into account in the single-particle distribution function. Here, we have assumed that the tracers do not interact among themselves. This is usually a very good approximation if the tracers are sufficiently dilute.


\paragraph{Isolated systems are governed by the Liouville equations, which preserves the Gibbs entropy. So the entropy of the system tracers + medium cannot decrease.} If that were true, then the entropy of the system tracers + medium would not increase in a normal (homogeneous) diffusion process, which is an absurd conclusion. This argument is a version of the irreversibility paradox, which plagues the whole of statistical mechanics (and not just diffusion phenomena); we do not wish to address this thorny issue here, and refer the reader to the vast literature on the subject (and in particular to \cite{Jaynes1957}).


\section{Conclusion}\label{conclusion}

We have defined and illustrated the concept of ``geometric ratchets''. Just like standard ratchets, these are based on the rectification of thermal fluctuations by an inhomogeneous background, and result in self-organization; the main difference with the latter is that geometric ratchets do \emph{not} involve net energy variations or heat transfers. We have argued that the geometric ratchet effect---which has actually been observed \cite{VanMilligen2005}---drives self-organization without heat exchanges: a phenomenon which is generally believed to be impossible.

It is not difficult to imagine applications of diffusion-driven traps such as the one discussed in sec. \ref{antidiffusion}. For instance, if a thermally-conducting material could be engineered in such a way that $(i)$ its thermal diffusivity has the above shape, and $(ii)$ the corresponding heat equation takes the FP form \eqref{FP}, then we have shown that  such a system would behave as a \emph{heat sink}. Similarly, we expect that light diffusion in a medium with continuously varying index (see \cite{DiFalco2011} for recent experimental developments in this direction) could result in a diffusive \emph{focusing effect}. 


We have also discussed the implications of these systems for the second law of thermodynamics: although they do not contradict any fundamental physical law, geometric ratchets do force us to reconsider statements such as ``the entropy of an isolated system can never decrease''. We have argued that, in general, these must be qualified by the reference to a suitable prior; without this caveat, they are simply not true.

One tantalizing question that we have not addressed in this work is whether useful work can be extracted from a thermal bath by means of the geometric ratchet mechanism. We leave this question open for further investigation, and close with Eddington's beautiful words \cite{Eddington2005}:




\begin{quote}
``If someone points out to you that your pet theory of the universe is in disagreement with Maxwell's equations---then so much the worse for Maxwell's equations. If it is found to be contradicted by observation---well these experimentalists do bungle things sometimes. But if your theory is found to be against the second law of thermodynamics I can give you no hope; there is nothing for it but to collapse in deepest humiliation.''
\end{quote}

\begin{center}
\line(1,0){100}
\end{center}

M. S. thanks M. Polettini for useful critical comments on a previous version of this manuscript, S. Carrozza and T. Koslowski for inspiring remarks on the relativity of entropy, H. Carteret for pointing out the similarity with ratchets, and H. Haggard for suggesting the name ``geometric ratchet''.

\bibliographystyle{utcaps}
\bibliography{library}

\end{document}